\documentstyle[emulateapj,psfig,flushrt,tighten, color]{article}

\def\msun{\,{\rm M_\odot}}

\newcommand\beq{\begin{equation}}
\newcommand\eeq{\end{equation}}
\newcommand{\ba}{\begin{eqnarray}}
\newcommand{\ea}{\end{eqnarray}}  
\def\spose#1{\hbox to 0pt{#1\hss}}
\def\lta{\mathrel{\spose{\lower 3pt\hbox{$\mathchar"218$}}
     \raise 2.0pt\hbox{$\mathchar"13C$}}}
\def\gta{\mathrel{\spose{\lower 3pt\hbox{$\mathchar"218$}}
     \raise 2.0pt\hbox{$\mathchar"13E$}}}

\righthead{Observing gravitational waves from the first black holes}
\makeatletter

\newenvironment{figurehere}
  {\def\@captype{figure}}
  {}
\makeatother
\begin{document}

\title{Observing gravitational waves from the first generation of black holes}

\author{Alberto Sesana\altaffilmark{1}, Jonathan Gair\altaffilmark{2}, Ilya Mandel\altaffilmark{3}\& Alberto Vecchio\altaffilmark{4}}
\altaffiltext{1}{Center for Gravitational Wave Physics, The Pennsylvania State University, University Park, PA16802, USA}
\altaffiltext{2}{Institute of Astronomy, University of Cambridge, Cambridge, CB3 0HA, UK}
\altaffiltext{3}{Department of Physics and Astronomy, Northwestern University, Evanston, IL, 60208 USA}
\altaffiltext{4}{School of Physics and Astronomy, University of Birmingham, Edgbaston, Birmingham, B15 2TT, UK}

\begin{abstract}
The properties of the first generation of black-hole seeds trace and distinguish different models of formation of cosmic structure in the high-redshift universe. The observational challenge lies in identifying black holes in the mass range $\sim 100 - 1000\,M_\odot$ at redshift $z\sim 10$.  The typical frequencies of gravitational waves produced by the coalescence of the first generation of light seed black-hole binaries fall in the gap between the spectral ranges of low-frequency space-borne detectors (e.g., LISA) and high-frequency ground-based detectors (e.g., LIGO, Virgo and GEO 600).  As such, these sources are targets for proposed third-generation ground-based instruments, such as the Einstein Telescope which is currently in design study.  Using galaxy merger trees and four different models of black hole accretion --- which are meant to illustrate the potential of this new type of source rather than to yield precise event-rate predictions --- we find that such detectors could observe a few to a few tens of seed black-hole merger events in three years and provide, possibly unique, information on the evolution of structure in the corresponding era.  We show further that a network of detectors may be able to measure the luminosity distance to sources to a precision of $\sim30\%$, allowing us to be confident of the high-redshift nature of the sources.  
\end{abstract}


\section{Introduction}

An ability to probe the nature of the first massive black-hole (MBH) seeds at
medium-to-high redshift
is fundamental to understanding the 
hierarchical assembly scenario, to discriminate between different models of
structure formation in the high-redshift universe and to explore the link between
black holes residing at the centre of galaxies and 
their hosts (Sesana, Volonteri \& 
Haardt 2007). 
If MBH seeds are \emph{massive} (i.e., $\sim 10^5\msun$), the future space-borne gravitational wave (GW) detector LISA (Bender et al. 1998) will probe the first epoch of mergers between these seeds at high redshift. However, if MBH seeds are \emph{light} (i.e., $\sim 100\msun$), the GWs from these mergers will fall between the sensitive frequency band of LISA and of currently operating and planned Advanced versions of ground-based instruments --- LIGO, Virgo and GEO 600 (see Sigg et al. 2008, Acernese et al. 2008 and Grote et al. 2008 for recent status reports). In this \emph{Letter} we show that third-generation laser interferometers may be able to fill this gap by directly probing the first mergers between light MBH seeds and will thus provide complementary information to other instruments.

If seed black holes (BHs) are the 
remnants of Pop III stars with mass $\sim 100\msun$ (Madau \& Rees 2001; see 
Sesana, Volonteri \& Haardt 2007 for a short review of this and alternative 
scenarios) we expect dozens of MBH binary (MBHB) coalescences per year in the mass range  
$\sim10^2-10^6\msun$ (Sesana et al.~ 2004) (NB we will use MBHB liberally to refer to any binary formed between black holes in the centres of merging dark-matter halos). 
Most of the MBHB events occur at redshift $z \gtrsim 3$, with the consequence that LISA will be able to detect MBHBs down to $\sim 10^3\msun$ with a signal-to-noise ratio (SNR) 
$\gtrsim10$, but will not probe lower masses. To probe the $10^2-10^3\msun$ range, a new GW telescope with optimal sensitivity in the frequency window $0.1-10$ Hz is needed. This could be achieved either by a second generation space-based detector, such as the Big Bang Observer (Phinney et al. 2003), ALIA (Bender et al. 2005) or DECIGO (Kawamura et al. 2006), or by the third generation of ground-based laser interferometers, for which the target is $\sim$ ten-fold strain sensitivity increase over Advanced LIGO and a low frequency cut-off at $\sim 1$ Hz (the ET target sensitivity is compared to other instruments in Fig.~\ref{f:shvsh}). 
In this \emph{Letter}, we focus on third-generation ground-based instruments as these are presently undergoing conceptual design studies, and for reference we use the specific example of the Einstein gravitational-wave Telescope (ET) (see e.g., Freise et al 2009).
%
We demonstrate that instruments such as the ET can detect seed black-hole binaries, albeit at likely rates of no more than a handful a year, and discuss whether such observations can uniquely identify these events as produced by light remnants of Pop III stars.

\section{Models for Pop III seed growth}

The astrophysical scenario that we consider in this \emph{Letter} 
assumes that seed black holes of mass $\sim 100\msun$ are produced by
the first generation of supernovae in the very high-redshift Universe 
at $z \approx 20$. These black holes are efficient at accreting mass 
and hierarchically merge following mergers between their host halos.
We can trace the merger hierarchy by means of Monte-Carlo merger-tree
realizations based on the extended Press-Schechter formalism (Press
\& Schechter 1974), assuming a standard $\Lambda$CDM
cosmology with the 1-year {\it WMAP} cosmological  parameters (Spergel et al.~2003); technical details are given in Volonteri, Haardt \& Madau (2003; VHM) and Volonteri, Salvaterra \& Haardt (2006; VSH).  
We consider four variants of this model, which have the same merger history for the dark-matters halos, but different prescriptions for the mass-distribution and accretion efficiency of the seeds:
(i) {\it VHM,ems} (VHM with equal-mass seeds) is based on equal $150\msun$ seeds accreting at the Eddington limit during each merger episode;
(ii) {\it VHM,smd} (VHM with seed-mass distribution) differs from {\it VHM,ems} only in the seed-mass distribution, which is log-normal in the range $10-600\msun$;  
(iii) {\it Shank} has black-hole seeds with flat mass distribution
in the range $150-600\msun$ and \emph{redshift-dependent} accretion efficiency, following
Shankar et al.~2004; 
(iv) {\it Hopk} again assumes black-hole seeds with flat mass-distribution
in the range $150-600\msun$ and accretion efficiency that is  {\it luminosity-dependent} according to the prescription given by Hopkins et al.~2005. 


Integrated over cosmic history, all these models reproduce the X-ray and optical quasar luminosity function at $z<3$, the observed faint X-ray counts of AGNs (see VSH), and, integrated over all MBH masses, predict about $50-70$ black-hole coalescences per year in the Universe. This range is statistical and does not include uncertainties in assumptions about, e.g., cosmology, that go into the merger trees, which could change the number of predicted events by a factor of a few in either direction. The accretion models considered here have been tuned to reproduce observations at low redshift, $z \lesssim 3$, rather than accretion onto seed black holes at {\bf $z \gtrsim 5$}. Recent work has indicated that accretion onto $100M_{\odot}$ black holes is inefficient (Alvarez et al. 2008, Milosavljevic et al. 2008) and generally sub-Eddington. 
This adds further uncertainties to the light seed scenario, but we emphasize that our choice of models is guided by the goal of illustrating the science potential of this new class of observations rather than an attempt to provide solid predictions for event rates.
%
%


\begin{figurehere}
\centerline{
\includegraphics[angle=0,width=3.2in]{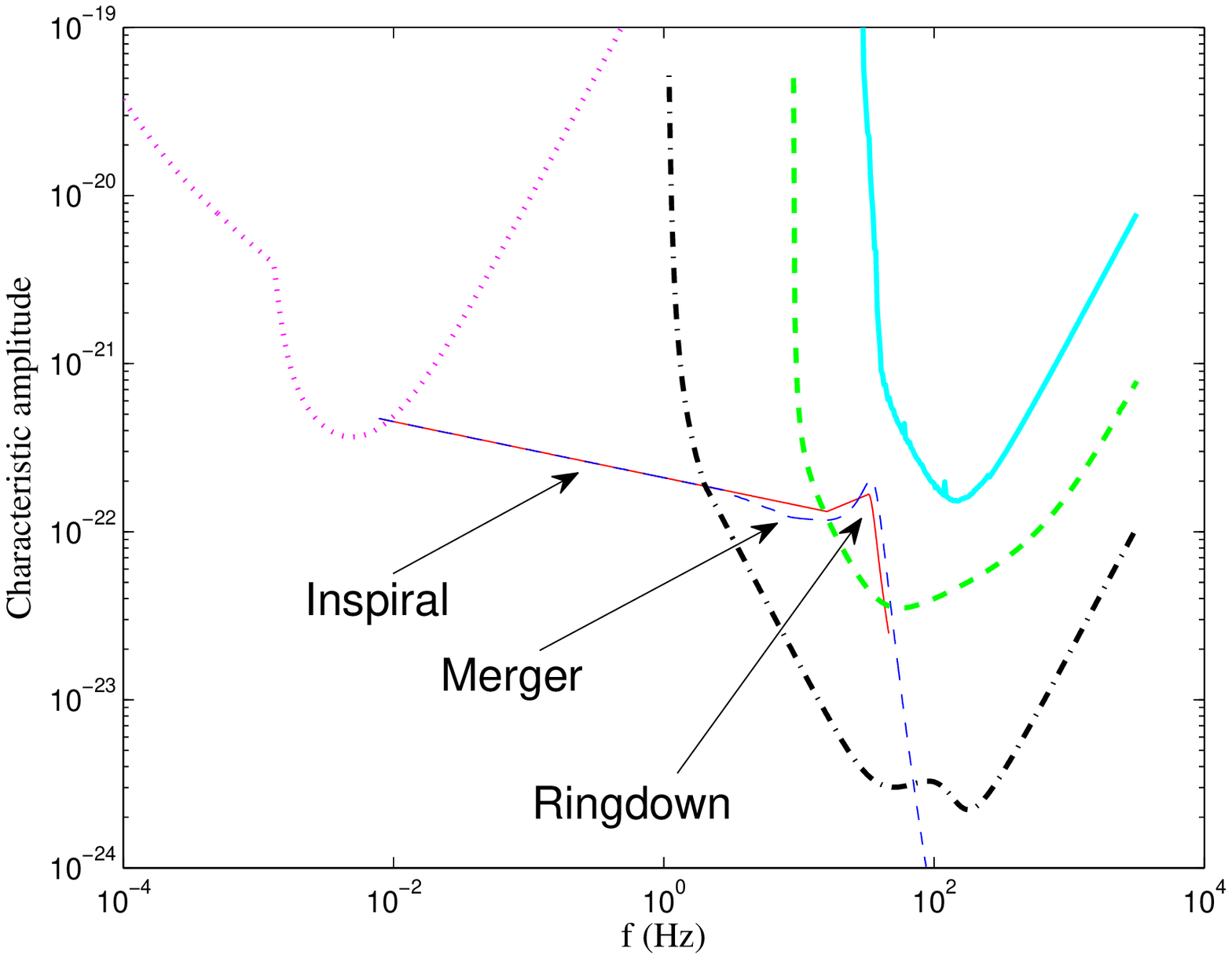}}
\caption{\footnotesize The thin curves show the characteristic amplitudes  $f \tilde{h}(f)$ of the frequency-domain gravitational waveforms IMR (solid red) and EOBNR (dashed blue) for an equal-mass, optimally-oriented source with a redshifted total mass of $500 M_\odot$ at a luminosity distance of $33$ Gpc.  The thick curves show the characteristic noise amplitude spectral densities $\sqrt{f S_n(f)}$ for LISA (dotted magenta), Enhanced LIGO (solid cyan), Advanced LIGO (dashed green) and ET (dash-dotted black).  The ratio of the characteristic amplitudes of the waveform to the noise yields the SNR accumulated in a bandwidth equal to the frequency.}
\label{f:shvsh}
\end{figurehere}

\section{Waveform and detector models}

To compute the sensitivity of the ET to GWs generated during the coalescence of seed MBH binaries, we model the gravitational wave emission with the phenomenological waveform family (IMR) introduced by Ajith et al. (2008). This describes radiation from a non-spinning black-hole binary and includes in a self-consistent manner the inspiral, merger and ring-down phases; these phases are marked in Fig.~\ref{f:shvsh}, which shows a typical frequency-domain gravitational waveform $\tilde h(f)=A(f) e^{i\psi(f)}$.  
Fig.~\ref{f:shvsh} also indicates that a significant fraction of the signal is contributed by the merger and ringdown, so an inspiral-only waveform would be inadequate.

Exact expressions for the amplitude $A(f)$ and the phase $\psi(f)$ of the IMR waveforms are provided by Eqs. (4.17), (4.18) and (4.19) and Table I of Ajith et al.~(2008). The strain at the detector depends on the total redshifted mass $M_z \equiv (1+z)M$, the symmetric mass ratio $\eta$, a fiducial time of arrival parameter $t_0$, and six extrinsic parameters --- two sky-position angles,  the orbital phase at time $t_0$, the wave polarization angle $\psi$, the source inclination angle with respect to the line of sight $\iota$, and the luminosity distance to the source $D_L$. The optimal signal-to-noise ratio (SNR) at the instrument is SNR$^2 = 4\int_0^\infty df |\tilde h(f)|^2/S_n (f)$, where $S_n (f)$ is the one-sided noise power spectral density of the interferometer, as shown in Fig.~\ref{f:shvsh}. For a network of detectors, the total network SNR is obtained by adding the SNRs of the individual instruments in quadrature.  We report SNRs averaged over the sky location of the source and its orbital-plane orientation. 

As a check, we also computed the SNR for some events using the effective-one-body-numerical-relativity (EOBNR) waveform family introduced by Buonanno et al.~(2007). The SNRs predicted by the EOBNR waveforms are typically somewhat higher (by up to $\sim25\%$) for equal-mass events, and are comparable for all events except those with very asymmetric mass ratios, $\eta \ll 1$, where neither waveform family has been shown to be valid.  

The currently favoured design for the Einstein Telescope calls for a 10km triangular facility containing three 60$^{\circ}$ detectors (Freise et al. 2009); we refer to this design as a ``single ET''. We use the noise power spectral density defined in Hild et al. (2008) for a single right-angle 10km detector. The angle-averaged sensitivities achieved by two right-angle detectors and three 60$^{\circ}$ detectors (single ET) are factors of $\sqrt{2}$ and $3/2$ higher, respectively, than for one right-angle 10km instrument.

\section{Detection-rate estimates}

For each of the four scenarios described in Section 2, we have generated 1000 independent realisations of the galaxy merger history and computed the SNR of all coalescences that take place over a time span of 3 years, representative of a typical data-taking period for a third-generation instrument. Figure \ref{f:nvssnr} shows the number of events that would be seen by the third generation network in three years as a function of the SNR in a single right-angle 10km detector within the network, for each of the four models. Assuming that a network SNR of $8$ would be required for detection, the SNR in one 10km detector would be 5.3 for a single ET or 4.4 (3.9, 3.1) if one (two, three) additional 10km detector(s) were added. 
Figure~\ref{f:nvssnr} indicates that in all models except {\it Shank}, we could expect to detect a few to a few tens of events over three years. The {\it Shank} model predicts fewer than one event in a single ET.

\begin{figurehere}
\vspace{0.5cm} 
\centerline{
\includegraphics[angle=270,width=2.8in]{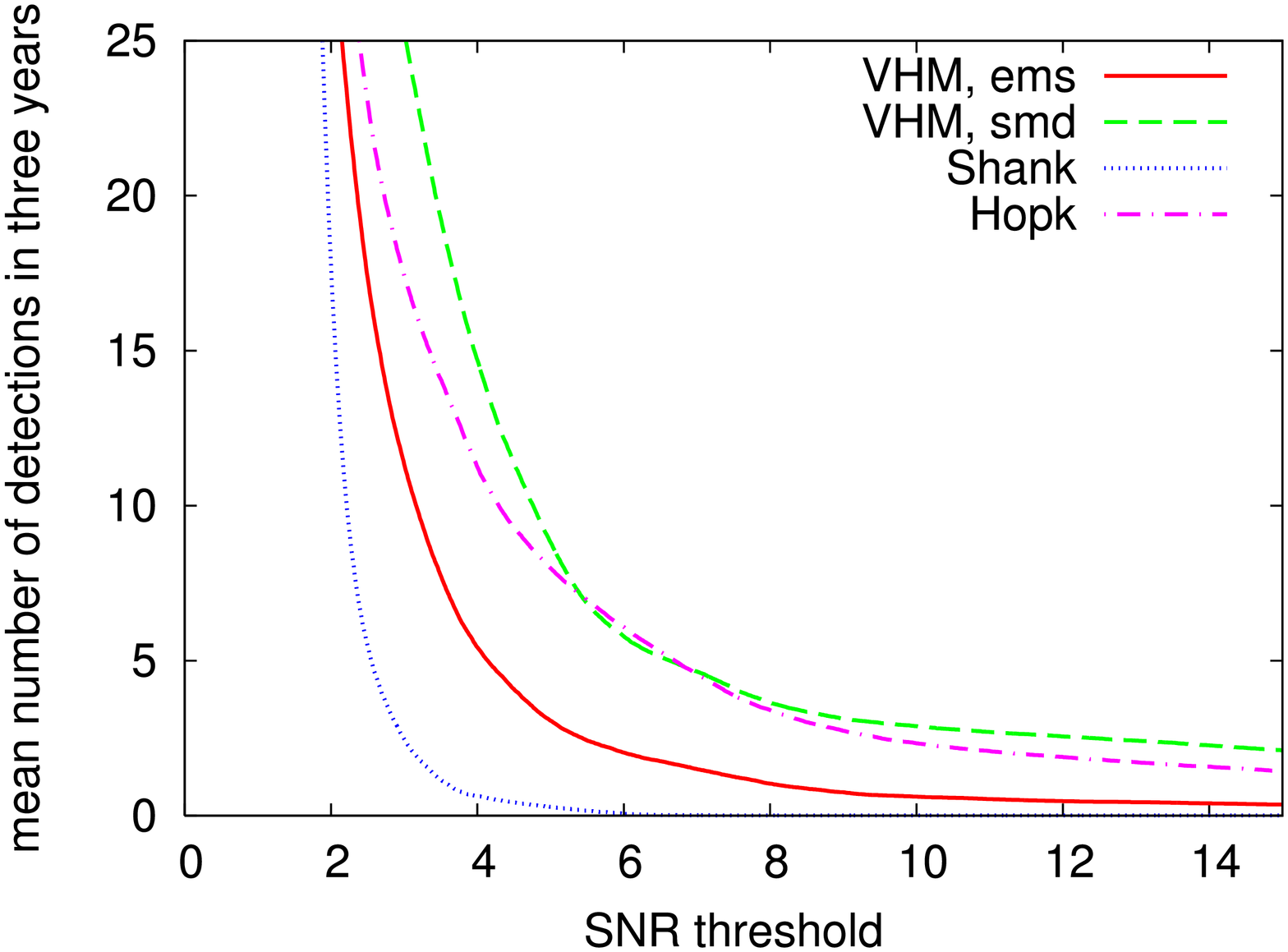}}
\vspace{-0.0cm}
\caption{\footnotesize Number of events detected in three years as a function of threshold SNR required in a single right-angle 10km detector. For a network SNR of 8, this threshold would be 5.3 for a single ET, or 4.4 (3.9, 3.1) with one (two, three) additional 10km detector(s).} 
\label{f:nvssnr}
\vspace{+0.5cm}
\end{figurehere}

The ET will be sited underground, which should make it possible to suppress seismic gravity-gradient noise (Hughes \& Thorne 1998) and achieve sensitivity at frequencies as low as $\sim 1$ Hz (cf.~Advanced LIGO, Fig.~\ref{f:shvsh}); however, it is not currently clear whether it will be possible to mitigate this noise source below $\sim 3$ Hz.  In addition, the ET may suffer a confusion background in the $1$--$5$Hz range due to cosmological compact-binary systems (Regimbau \& Hughes 2009). The impact of such a background on Pop III event rates needs to be properly quantified, but we have checked this crudely. The preceding results assumed a low-frequency cut-off of $1$Hz (which is also the cut-off we use for parameter estimation), but we found that only $\sim 25\%$ of the events were lost
with a higher cut-off at $5$Hz. We conclude that design changes or a confusion background should not affect our qualitative  conclusions.

Figure~\ref{f:nvsMz} indicates the distribution of masses and redshifts for the events that would be detected under each scenario. We see that most events have intrinsic masses of a few hundred $M_\odot$ (down to a few tens in the {\it VHM,smd} model), although a few events with $M \gtrsim 1000\msun$ might also be observed. Many of the events will be at medium-to-high redshift, $5\lesssim z \lesssim 10$, except in the {\it VHM,smd} model which predicts a distribution peaked around $z \sim 4$ with a long tail extending to $z>12$.
Most of the observable events are related to the hierarchical assembly of small--to--medium size parent halos (masses in the range $10^{11}-10^{13}\msun$ at $z=0$). In such halos, seeds are less likely to experience coalescences accompanied by major accretion episodes at high redshift, and they are more likely to have $M \lesssim1000\msun$ at $z=10$, leaving them in the band accessible to third generation detectors. 

The number of events and their redshift distribution may provide constraints on both the mass function and accretion history of seed black holes in the early Universe.  Considering our example scenarios, in the {\it Shank} model seeds are born quite massive ($150-600\msun$) and they accrete at the Eddington limit at high redshift [see equation (2) in VSH]; in this model, most of the seeds grow above $1000\msun$ by $z=10$, making detection with ground-based detectors difficult. 
In the {\it Hopk} model, accretion is on average less efficient and many of the seeds are still in the few-hundred solar-mass range at $z<10$, making detection easier. The initial seed mass distribution may also leave a detectable signature on the observed events. The {\it VHM,ems} and {\it VHM,smd} models are characterised by the same Eddington-limited accretion prescription; nonetheless, their mass and redshift distributions are significantly different. In the {\it VHM,ems} model most of the $150\msun$ seeds grow at high $z$, radiating outside the $1$--$1000$Hz band; only the `tail' of $\lesssim 10^3\msun$ black holes left behind at $z<10$ is detectable. In the {\it VHM,smd} model, there are many $<100\msun$ seeds which would be observable at very high redshift ($z>10$). Many of them will grow inefficiently, and still have masses $\sim100\msun$ at low redshift, making them perfect targets for ET; however, it may be impossible to discriminate between coalescences involving these low-$z$ Pop III remnants and those involving intermediate-mass BHs (IMBHs) formed at low redshift. IMBHs might form via runaway collisions of massive stars in globular clusters (Portegies-Zwart et al.~1999; but see Glebbeek et al.~2009), and scenarios have been proposed in which IMBH binaries also form (Fregeau et al. 2006). The IMBH masses considered there, $\sim 10^3M_{\odot}$, are somewhat higher than seed black hole masses, $\sim$few$\times100M_{\odot}$. Therefore, if we observe an IMBH merger at $z\gtrsim 5$ with masses $\sim100M_{\odot}$, we can be fairly confident that the constituent black holes are Pop III seeds. 
Further work will be needed to identify what observational signature(s) provide the best discriminating power between these two formation channels.

\begin{figurehere}
\vspace{0.5cm} 
\centerline{
\includegraphics[width=3.0in]{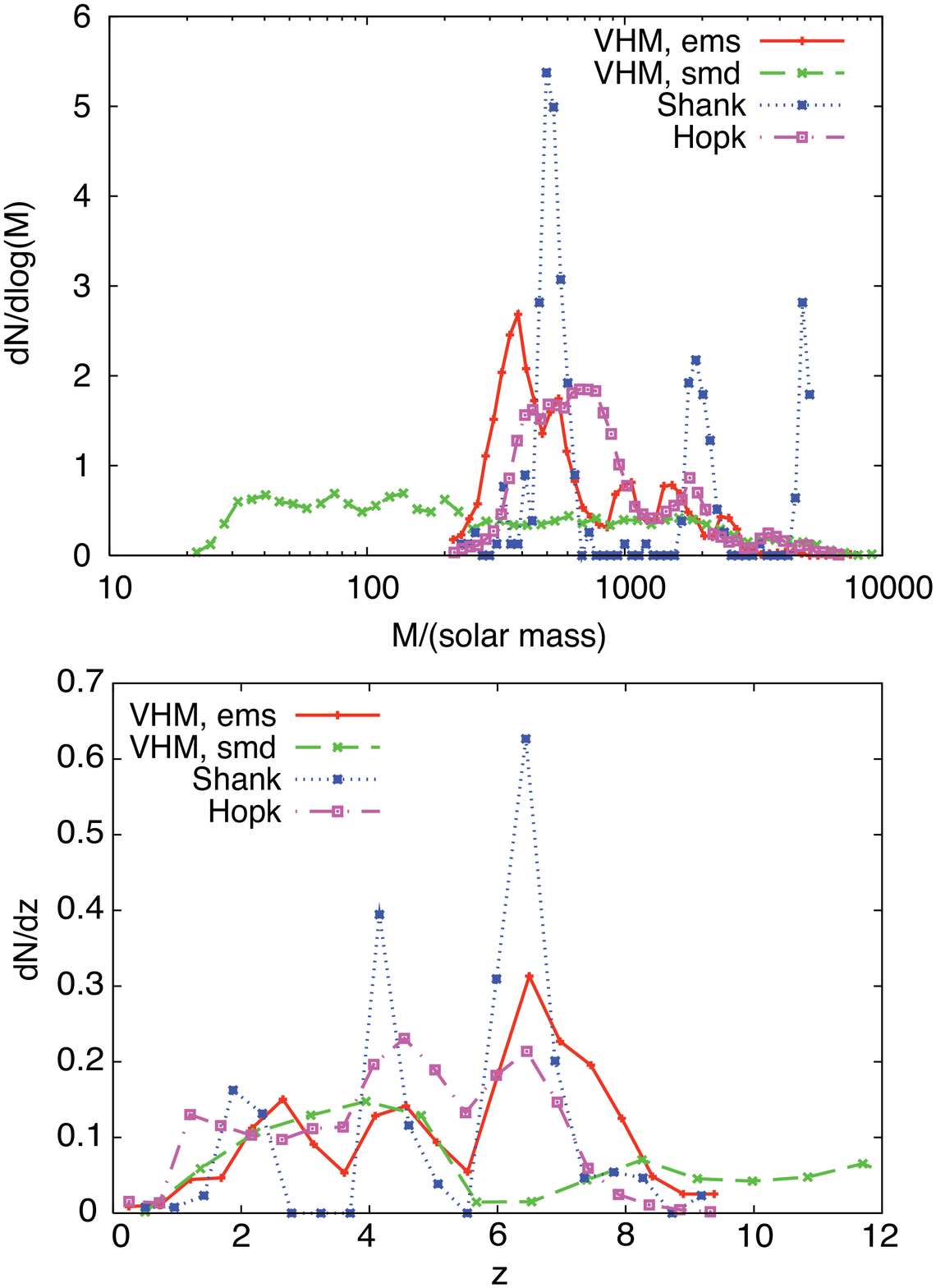}}
\vspace{-0.0cm}
\caption{\footnotesize Distribution of source-frame masses (top) and redshifts (bottom) for detected (SNR$\ge 5$) events, normalised such that the integral over the distribution is unity.}
\label{f:nvsMz}
\vspace{+0.5cm}
\end{figurehere}

This current work indicates that, while probing seeds mostly at $z<10$, third-generation detectors may provide useful insights into the properties and accretion history of seed black holes at higher redshift.
LISA will be able to probe deeper in redshift, since ground-based detectors are limited at low frequency by a gravity-gradient `noise wall' at 1Hz.  However, LISA will not be able to observe seed black-hole mergers because it is not sufficiently sensitive in the relevant frequency range (see Fig.~\ref{f:shvsh}).  LISA will only provide indirect constraints on seed black-hole populations through observations of subsequent mergers of black holes in the mass-range $10^4$--$10^6\ M_\odot$.

\section{Parameter-estimation results}
In order to identify an event as a seed black-hole merger, we must be able to determine the redshift and mass of the system. As mentioned earlier, events at low redshift $z \lesssim 3$ might not involve primordial black holes. A gravitational-wave observation determines the redshifted total mass $(1+z)\,M$ and luminosity distance $D_L(z)$, but not the redshift $z$ independently. It is unlikely that electromagnetic counterparts to ET events will be observed, but if we assume a concordance cosmology inferred from other observations, we can determine the redshift from the luminosity distance. The fractional redshift error is then comparable to the fractional luminosity-distance error, plus an $O(10\%)$ error from uncertainties in the cosmology.

We used the Fisher information matrix to evaluate the parameter-estimation errors.   We carried out a Monte Carlo simulation over the extrinsic parameters for fixed choices of the intrinsic parameters, $M_z$ and $\eta$.  We note that this approach may overstate the precision of parameter estimation at low SNRs (Vallisneri 2008); a more rigorous study of parameter-estimation accuracy would require the computation of the full posterior probability density function and is beyond the scope of the present paper.

The intrinsic parameters, $M_z$ and $\eta$, are determined by the waveform phase evolution and so we expect to measure them well. The extrinsic parameters, by contrast, affect only the relative amplitudes of the signal. A single ET makes four independent measurements (two quadratures in two detectors) and so at least one additional interferometer is needed to determine the six extrinsic parameters.  We computed parameter-estimation errors for several network configurations consisting of an ET at the location of Virgo and additional interferometers at Hanford and Livingston. At a fixed network SNR of $8$, we found that for all networks $M_z$ and $\eta$ would be determined to a  fractional accuracy better than $1\%$ for all but the most massive systems ($M_z \sim 1000 M_\odot$).
The distance, and hence redshift, will be determined less accurately. One ET plus a second 10km interferometer will be enough to determine the distance to better than $\sim40\%$ in $68\%$ of cases. The addition of a third 10km detector will improve this distance precision to $\sim30\%$. Upgrading the 10km detectors to ETs leads to further modest improvements, but would also increase the SNR accumulated from a source at a fixed distance. We also explored siting the second detector in Perth, Australia, instead of Hanford, but found little net difference in parameter-estimation accuracy.  
The error in the source-frame mass is dominated by the error in $z$ rather than that in $M_z$, and so is also $\sim40\%$. 
Therefore, we should be able to say confidently that a source at $z\sim 5$ \emph{is} at high redshift  and has $M\sim 100\msun$, and hence is most likely a seed black-hole merger.

\section{Discussion} 
Information about seed black holes is extremely difficult to
obtain using present or future electromagnetic observations. The only direct means to study seed black holes is via gravitational-wave observations of mergers. If seed black holes are remnants of Pop III stars at high redshift, then LISA may not be able to probe the early stages of their evolution.
In this paper we have analyzed the ability of third-generation ground-based 
interferometers, such as the proposed Einstein Telescope, to detect the coalescences of
$100-1000\msun$ seed black-hole binaries for a range of seed 
properties and accretion histories. 

We have found that third-generation detectors will be able to detect $\sim1$--$30$ 
events over a 3-year observation, depending on the selected model and on 
the assumed telescope configuration. The distribution of detected masses, 
ranging from a few$\times10\msun$ to a few$\times 1000\msun$, is complementary to the
range probed by LISA, making the detection of low-mass seeds
possible. The noise wall at 1 Hz will preclude the detection of sources
at very high redshift, although in the case of a seed-mass spectrum that extends down to $10\msun$ (i.e., the {\it VHM,smd} model), some detections may be possible at $z>10$. The mass and redshift distribution of detected events may be useful in reconstructing the accretion history of the first
seeds. We have also shown that a detector network with at least two sites will be able to determine the (redshifted) masses and luminosity distances of the majority of events to accuracies of 
$\lesssim1\%$ and $\lesssim 40\%$ respectively. This should be sufficient for us to say with confidence that the merger is occurring at high redshift between low-mass seeds.

Previous work (Wyithe \& Loeb 2004) indicated that Adv. LIGO might detect pop III seeds. That work used a semi-analytic model for structure growth and our current work contradicts this, suggesting that their model significantly over-predicted the number of mergers occurring at low redshift. There are various sources of uncertainty in our analysis. Our waveform model ignored the effects of black-hole spin and higher gravitational-wave harmonics, which tend to enhance the SNR and increase the event rate. These effects also help to break degeneracies between parameters, improving the validity of the Fisher Matrix approach to computing parameter accuracies used in this paper. However, the SNR threshold of $8$ may be optimistic when source confusion and realistic instrumental noise 
are taken into account. The merger-tree models have various uncertainties, such as the choice of cosmological parameters, e.g., $\sigma_8$, which could change the rates by a factor of a few in either direction. More work is also needed to understand how to distinguish IMBH binaries formed in globular clusters from seed black holes. Despite these uncertainties, this {\it Letter} is a proof of concept which clearly demonstrates that third-generation ground-based instruments have the capability to detect seed black holes, allowing us to confirm (or discard) the hypothesis that MBH seeds are light remnants of Pop III stars, and suggests that this could be one of the science drivers for these instruments.

\section*{ACKNOWLEDGMENTS}
We would like to thank M. Volonteri for useful discussions and for providing the Monte-Carlo realisations of the halo and MBH merger hierarchy. AS acknowledges support from 
NSF Grant No.s~PHY 06-53462 and PHY 05-55615, and NASA Grant No.~NNG05GF71G, awarded to The Pennsylvania State University.  AV is partially supported by the UK Science and Technology Facilities Council. JG's work is supported by the Royal Society. IM is partially supported by NASA ATP Grant NNX07AH22G to Northwestern University. 

{}
\end{document}